\documentclass[twocolumn,prl,showpacs,noshowkeys]{revtex4-1}

\usepackage{amsfonts}
\usepackage{amssymb}
\usepackage{graphicx}
\usepackage{dcolumn}
\usepackage{bm}
\usepackage{ulem}
\usepackage{color}

\begin{document}

\title{Prediction of weak topological insulators in layered semiconductors} 
\author{Binghai Yan$^{1}$, Lukas M\"{u}chler$^{1,2}$, Claudia Felser$^{1,2}$}
\affiliation{$^1$ Institute for Inorganic and Analytical Chemistry, Johannes Gutenberg University of Mainz, 55099 Mainz, Germany \\ 
$^2$ Max Planck Institute for Chemical Physics of Solids, D-01187 Dresden, Germany}

\date{\today}

\begin{abstract}
We report the discovery of  weak topological insulators by \textit{ab initio} calculations in a honeycomb lattice. We propose a structure with an odd number of layers in the primitive unit-cell as a prerequisite for forming weak topological insulators. Here,  the single-layered KHgSb is the most suitable candidate for its large bulk energy gap of 0.24 eV. Its side surface hosts metallic surface states, forming two anisotropic Dirac cones. Though the stacking of even-layered structures leads to trivial insulators, the structures can host a quantum spin Hall layer with a large bulk gap, if an additional single layer exists as a stacking fault in the crystal. The reported honeycomb compounds can serve as prototypes to aid in the finding of new weak topological insulators in layered small-gap semiconductors.

\end{abstract}
\maketitle

Topological insulators (TIs) have attracted extensive research attention in the last few years ~\cite{qi2010,moore2010,hasan2010,qi2011RMP}.  They differ from normal insulators in that they exhibit gapless boundary states inside a bulk energy gap. In two dimensions (2D), a TI has counter-propagating edge states with opposite spin that carry dissipationless current~\cite{bernevig2006d,koenig2007}. In three dimensions (3D), TIs are further classified into strong and weak TIs according to the nature of their surface states~\cite{fu2007a,moore2007,fu2007b}. A strong TI has metallic surface states that usually form an odd number of Dirac cones. The surface states are robust to perturbations that do not break the time-reversal (TR) symmetry. A weak TI is topologically equivalent to a stack of 2D TI layers; however such a TI has an even number of Dirac cones on the side surface. Its surface states were initially assumed to be unstable with respect to disorders~\cite{fu2007a}; however recent research~\cite{ringel2011,mong2012} has revealed that surface states of a weak TI behave robust even under strong TR invarant disorders, similar to those of a strong TI.

The discovery of TI materials has stimulated intense research activity. The first experimentally observed TI was HgTe~\cite{bernevig2006d,koenig2007}. This compounds exhibits an inverted band order between the conduction and valence bands (Hg-$s$ and Te-$p$ states), which determines its band topology. This intuitive band-inversion picture with the topological band theory~\cite{kane2005A,fu2006,moore2007,roy2009b} paved the way to the discovery of new TIs including the Bi$_2$Se$_3$ family~\cite{zhang2009,xia2009,chen2009}, the Heusler family~\cite{chadov2010,lin2010a}, and the TlBiSe$_2$ family~\cite{yan2010,sato2010,kuroda2010,chen2010,lin2010b}. Nearly all of reported TIs are either 2D TIs or 3D strong TIs; however, no weak TIs have been reported thus far. In this Letter, we report a method to design weak TIs using layered semiconductors. We employed the honeycomb lattice as an example. Several ternary compounds in the double-layered honeycomb lattice 
have previously been  reported to be trivial insulators~\cite{zhang2011}, although they exhibit band inversions. In our study, we used an odd number of layers in the primitive unit-cell in order to realize an odd number of band inversions. Using this approach we discovered a family of weak TIs among ternary honeycomb compounds. Taking advantage of the tunability of the band structures of these compounds, we demonstrated a phase-diagram including weak TIs, strong TIs and trivial insulators.

\begin{figure}
    \includegraphics[width=3.5 in]{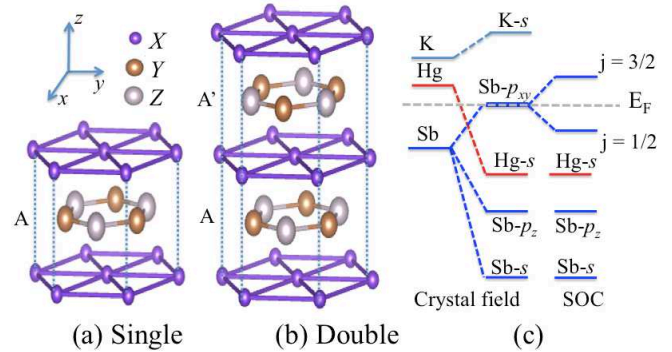}
    \caption{Crystal structures of  honeycomb compound $XYZ$. The single-layered structure (a) contains only one $XYZ$ layer, while the double-layered one (b) includes two $XYZ$ layers.  These two $XYZ$ layers in (b) are labeled as A and A$^\prime$, which are related to each other by an inversion with one $X$ atom as the inversion center.   (c) The band evolution at the $\Gamma$ point of the single-layered KHgSb. Under crystal filed without SOC,  Sb-$p_{xy}$ states are degenerate at the Fermi energy and the Hg-$s$ state is below these degenerate states. After applying SOC, Sb-$p_{xy}$ splits into $j = 3/2$ and $j = 1/2$ states, thereby resulting in a gapped insulator with inverted bands between Hg-$s$ and Sb-$p_{xy}$($j = 3/2$).}
\end{figure}

The $XYZ$ honeycomb compound can typically be viewed as honeycomb $YZ$ layers with alternating hexagonal layers of $X$ atoms stuffing between neighboring $YZ$ layers. This is similar to the cubic $XYZ$ Heulser compound (see ref.\onlinecite{graf2011} and references therein), in which $Y$ and $Z$ atoms form a zinc blende structure with the $X$ atoms filling the void space of the lattice. The single-layered lattice has only one honeycomb layer in the primitive unit cell, and it exhibits no inversion symmetry. However, the double-layered lattice contains two honeycomb layers with a formula of $X_2Y_2Z_2$, in which two $YZ$ layers are connected by the space inversion at $X$, as shown in Fig.~1. The low energy band structure is mainly related to the $YZ$ honeycomb layer, while the $X$ layer affects the coupling between the $YZ$ layers. It is possible to create a $YZ$ layer with an inverted band structure using heavy elements that have strong spin-orbit coupling (SOC). Such a layer leads to the formation of a 2D TI, also called a quantum spin Hall (QSH) insulator. By stacking such $YZ$ layers along the $z$ direction, we can obtain weak TIs~\cite{fu2007a} by retaining an odd number of layers in the primitive unit-cell. Here the odd-layered stacking induces an odd number of band inversions, which is necessary to realize topological nontrivial band structures. Moreover, it is also possible to tune the intra-layer band inversion and the inter-layer coupling by using different $X$, $Y$ or $Z$ elements. As a result, we can potentially realize the transition from weak to strong TIs, and even to trivial insulators.

In order to investigate band structures of the honeycomb compounds, we performed \textit{ab initio} calculations within the framework of the density-functional theory (DFT). The exchange-correlation functional was within the generalized gradient approximation~\cite{perdew1996}. The core electrons were represented by the projector-augmented-wave potential. We employed the \textit{Vienna ab initio simulation package} with a plane wave basis~\cite{kresse1993}.  First, we relaxed the lattice parameters and atomic positions for ternary compounds $XYZ$ ($X$ = K, Na, Li; $Y$ = Hg, Cd / Au, Ag; $Z$ = Sb, As, P / Te, Se). Here $XYZ$ follows the 18-valence-electron rule (closed shell), in a manner similar to the Heusler compounds~\cite{chadov2010,graf2011}. Subsequently, we calculated their band structures with SOC using these optimized structures.

We take KHgSb as an example of these compounds. In an isolated KHgSb single layer, (HgSb)$^-$ forms a honeycomb layer, while K loses one electron with the K$^+$ state considerably above the Fermi energy ($E_F$). The valence bands are composed of the Hg-$s$ and Sb-$sp$ states. As illustrated in Fig.~1c, the Sb-$s$ state is the bottom-most level, and the Hg-$s$ and Sb-$p$ states are located near $E_F$, in which Hg-$s$ is lower than Sb-$p$. Under the crystal field, Sb-$p$ splits into the $p_{xy}$ and $p_z$ states with $p_{xy}$ being degenerate at $E_F$. When applying SOC, the $p_{xy}$ states split again into $j = 3/2$ and $j = 1/2$ states, thereby causing the formation of a finite energy gap. Then we can see a band inversion between Hg-$s$ ($j = 1/2$) and Sb-$p_{xy}$ ($j = 3/2$) states, which occurs only at the $\Gamma$ point of the 2D Brillouin zone. And this inversion is confirmed by our DFT calculations. This results in a nontrivial topological $\mathbb{Z}_2$ index of $\nu$ = 1, according to the $\mathbb{Z}_2$ classification~\cite{fu2007a,fu2007b}. Therefore, an isolated KHgSb layer is a 2D TI with an inverted band structure, similar to HgTe quantum wells.  Next we stack the isolated KHgSb layer along the $z$ direction (perpendicular to the surface) into a 3D lattice. If a primitive unit-cell contains only one KHgSb unit, we refer to it as A-A stacking, where A denotes a single KHgSb layer.  If the unit-cell contains two KHgSb units, we refer to it as AA$^\prime$-AA$^\prime$ stacking, where A$^\prime$ represents an inverted A layer with the K atom as the inversion center. A-A stacking of KHgSb is the simplest case. Here, the inter-layer coupling is weak. Consequently, the band dispersion along the $\Gamma$-$Z$ (also called $A$) direction is exteremely small. Hence, the band inversion remains at both the $\Gamma$ and $Z$ points. As shown in Fig.~2a, the Hg-$s$ and Sb-$p_{xy}$ band inversion happens twice with one inversion occuring at the $\Gamma$ point and the other occuring at the $Z$ point. Thus, the single-layered KHgSb belongs to the $\mathbb{Z}_2$ (0;001) class of weak TIs. Here [001] also indicates the stacking direction of the QSH layers. The energy gap at $\Gamma$, which is approximately 0.5 eV, is determined by the SOC splitting of the Sb-$p$ states. The $s$-$p$ band inversion strength at $\Gamma$, which is approximately 1.5 eV, is related to the energy level difference between the Hg-$s$ and Sb-$p$ states. 

\begin{figure}
    \includegraphics[width=3.5 in]{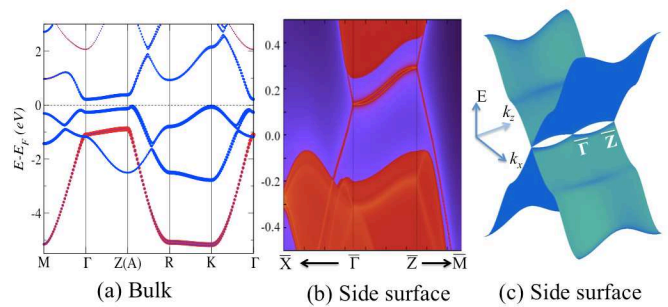}
    \caption{(a) Bulk band structures of single-layered KHgSb. The Hg-$s$ state is highlighted by filled red circles and Sb-$p$ states by filled blue circles. The size of circles indicates the amount of corresponding $s$ or $p$ character. The surface states on the side surface ($xz$) of KHgSb were calculated (b) by the tight-binding Wannier function method and (c) by the k$\cdot$p method. Two Dirac cones exist at the $\bar{\Gamma}$ and $\bar{Z}$ points, respectively.  
  }
\end{figure}

The QSH insulator exhibits gapless edge states forming a single Dirac cone at the $\Gamma$ point. When a 3D weak TI is formed by stacking QSH layers, these edge states interact with each other on the side surface due to inter-layer coupling. Consequently, most of these metallic states become gapped, while the Dirac point persists at the time-reversal-invariant-momenta (TRIM) as explained by the Kramer’s theorem. This results in an even number of surface Dirac cones. In order to uncover the surface states, we performed surface band structure calculations using maximally localized Wannier functions~\cite{marzari1997} extracted from our \textit{ab initio} calculations. In order to investigate a single surface, we applied the standard Green-function iteration method and obtained the density of states of a half-infinite surface, shown in Fig.~2b. The surface is selected as the side surface $xz$ of the KHgSb lattice, where the HgSb honeycomb layer has a zigzag type of termination on the boundary. Two Dirac points are observed at the $\bar{\Gamma}$ and $\bar{Z}$ points, which are TRIM. The surface states exhibit strong anisotropy, in which the energy dispersion is considerably larger along $k_x$ than that along $k_z$.   

As seen above, the 2D KHgSb layer has an $s-p$ band inversion similar to that of HgTe. Bernevig, Hughes, and Zhang (BHZ)~\cite{bernevig2006d} has previously written a $k \cdot p$ Hamiltonian for HgTe quantum wells. Hence, we can describe this QSH layer by the BHZ model. When stacking the QSH layers into 3D, we added the $k_z$ term to the BHZ model and obtained the Hamiltonian for our weak TIs, expressed on the basis of $\{|s; j=1/2, m_j = 1/2>$,  $|p; j = 3/2, m_j = 3/2>$, $|s; j = 1/2, m_j = -1/2>$, $|p; j = 3/2, m_j = -3/2>\}$,
\begin{eqnarray}
H(k)=\epsilon(k)\mathbb{I}_{4\times 4}+\left(\begin{array}{cccc}M(k)&A k_+& 0 & Ek_z\\
A k_-&-M(k)&Fk_z&0\\
0&Fk_z&M(k)&-A k_-\\
Ek_z&0&-A k_+ &-M(k)\end{array}\right),\label{eq:BHZ}
\end{eqnarray}\noindent
where $E(k)=C+Dk^2$ and $M(k)=M-B(k_x^2+k_y^2)-Gk_z^2$. We used a minimal four-band model, and the split-orbit split-off bands ($j=1/2$) are neglected, for they are not involved into the band inversion. Here, $M$ in $M(k)$ denotes the intra-layer band inversion strength, and $Gk_z^2$, $Ek_z$, and $Fk_z$ are related to the inter-layer interaction. This $k \cdot p$ model holds for both the $\Gamma$ and $Z$ points due to weak $k_z$ dispersion. As a simple example, eq. (1) represents a QSH layer when $C=D=0, A=1, B=-1, M=-2$, and $G=E=F=0$. The compound becomes a weak TI when we initiate the inter-layer coupling by setting $G=-0.1, E=F=0.1$. We subsequently transferred the above continuum model to a tight-binding lattice and solved the surface band structure on a slab configuration. The anisotropic surface states are displayed on the entire 2D Brillouin zone, showing two Dirac points similar to those obtained from our \textit{ab initio} calculations (see Figs ~2b and 2c). 

\begin{figure}
    \includegraphics[width=3.5 in]{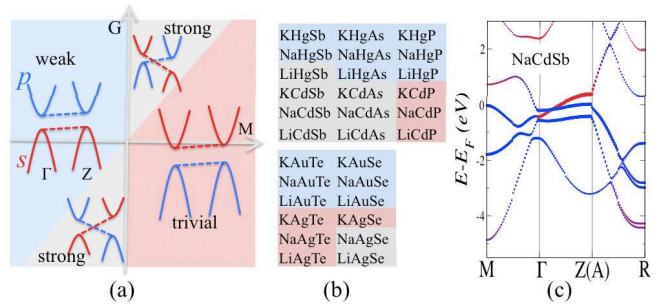}
    \caption{(a) Phase diagram with respect to band inverstion strength $M$ and inter-layer coupling $G$. The band inversions at $\Gamma$ and $Z$ are illustrated for the single-layered lattice. The dashed lines denote the band dispersion along $\Gamma-Z$ direction. The red (blue) solid lines indicate the $s$ ($p$)band dispersion in $k_x$$k_y$ plane around the $\Gamma$ and $Z$ points.  (b)  All single-layered XYZ compounds in this family are classified into weak TIs (light blue), strong TIs (gray), and trivial insulators (pink). (c) The band structure of single-layered bulk NaCdSb is shown as an example of a strong TI. The band inversion between the $s$ (red dots) and $p$(blue dots) states occurs at $\Gamma$, but disappears at $Z$.}
\end{figure}

Above model provides us a tool to investigate the phase diagram of the interplay between the band inversion strength (characterized by $M$) and the inter-layer coupling (mainly characterized by $G$), as illustrated in Fig.~3a. Hence, the condiction $M<0$ ($M/B>0$) represents a weak TI for weak inter-layer coupling ($G<0$). Moreover, we can still obtain a weak TI even when $G>0$, since the band inversions do not change for $\Gamma$ and $Z$. On the other hand, strong inter-layer coupling ($G<<0$) may remove the $s-p$ inversion at $Z$, giving rise to a strong TI. In general, the condiction $M>0$ represents a trivial insulator. However, the presence of very strong inter-layer coupling ($G>>0$) can possibly invert the bands at the $Z$ point, thereby resulting in a strong TI, too. Therefore, we can find strong TIs in honeycomb-lattice compounds, other than weak TIs and trivial insulators. The band structure of NaCdSb is shown in Fig.~3c as an example. The band inversion appears at the $\Gamma$ point, while it disappears at $Z$ because of considerable $k_z$ dispersion of the $s$-band. However, the system becomes a topological semimetal, instead of forming a real insulator. Here, the presence of three-fold rotational symmetry induces the band crossing between $\Gamma$ and $Z$, and an energy gap can form when this symmetry is broken (e.g. by in-plane strain)~\cite{zhang2011}.  We listed all the materials of the single-layered lattice in the phase diagram according to our \textit{ab initio} calculations. As shown in Fig.~3b, the heaviest compounds are usually weak TIs ($M<0, G<0$); the lightest ones are trivial insulators ($M>0$); and in between them are strong TIs ($M<0,G<<0$). In $X$Ag$Z$ compounds, strong TIs exist among the lightest compounds. This is because the Ag-$d$ states hybridize with Ag-$s$ and $Z-p$ states and violate the simple $s-p$ picture slightly. As mentioned previously all strong TIs are semimetals. Most of weak TIs are also found to be semimetallic, though they usually have a finite direct energy gap throughout the Brillouin zone. Among weak TIs, KHgSb and KHgAs have full energy gaps of 0.24 and 0.05 eV, respectively.

\begin{figure}
    \includegraphics[width=3.5 in]{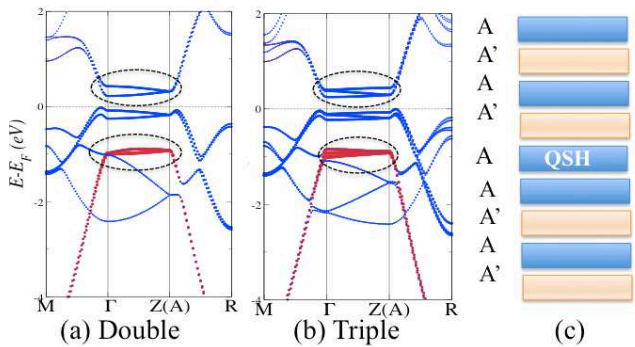}
    \caption{Bulk band structures of (a) double-layered KHgSb and (b) triple-layered KHgSb. Filled red circles denote Hg-$s$ states while blue circles indicate Sb-$p$ states.  The inverted Hg-$s$ and Sb-$p$ bands are highlighted by dashed circles. Band inversions occur twice and thrice in (a) and (b), respectively.  (c) The stacking fault in the honeycomb lattice. In a trivial insulator with AA$^\prime$-AA$^\prime$ stacking, an extra A layer as an stacking fault can become a quantum spin Hall layer.
    }
\end{figure}

The single-layered honeycomb lattice of KHgSb exhibits $s-p$ band inversions once at $\Gamma$ and once at $Z$, as discussed above. This picture can be generalized to multi-layered structures within the weak coupling limit. As indicated by the calculated band structures in Fig.~4a, the double-layered structure with AA$^\prime$-AA$^\prime$ stacking has band inversions twice at $\Gamma$ and twice at $Z$. Here, two Sb-$p$ bands exist at the conduction band bottom, while two Hg-$s$ bands appear in the valence bands. Thus, the double-layered KHgSb is a trivial insulator with $\mathbb{Z}_2$ (0;000), consistent with our previous results obtained from the parity criteria~\cite{zhang2011}. Furthermore, the triple-layered structure with AA$^\prime$A-AA$^\prime$A stacking is a weak TI again, due to inversion ocurring thrice (see Fig.~4b) at $\Gamma$ and $Z$, respectively. Therefore, we can expect an oscillation transition from a trivial insulator to a weak TI, when the primitive unit-cell changes from the even to odd-layered stacking. This even-odd transition is also consistent with the results of recent phenomenological study by Ringel \textit{et al.} ~\cite{ringel2011} and Mong \textit{et al.} ~\cite{mong2012}. If we treat the double(even)-layered structure as a period-doubling reconstruction~\cite{ringel2011,mong2012} of a single(odd)-layered lattice, the scattering between two original Dirac cones will localize the surface states and induce a gapped trivial surface. We can further estimate the stability of even and odd-layered structures via total energy calculations. Our DFT calculations show that  the double-layered lattice has the lowest energy, while the energy difference between single, double and triple-layered structures is below 0.05 eV per layer. Since there is no strong energetical preference of even or odd layers, it is possible to synthesize single-crystals of multi-layered structures by controlling experiment condicitons. On the other hand, in realistic materials the existence of disorders in the honeycomb plance can possibly eliminate the difference between A and A$^\prime$ layers, thereby giving rise to an equivalent A-A stacking lattice. Therefore, strong disorders in layered structures may probably induce weak TIs, rather than distroy them.

In the weak TI a one-dimensional helical state has previously been proposed to exist along the dislocation line of the lattice, which is protected by TR symmetry and stable against the weak disorder~\cite{ran2009,imura2011}. Our layered honeycomb compounds can provide an ideal platform to realize this proposal when dislocation defects exist. Another interesting topological defect is the stacking fault in the layered structure (illustrated in Fig.~4c). For example, in a double-layered trivial insulator a stacking fault occurs in the sequence of AA$^\prime$-AA$^\prime$-A-AA$^\prime$-AA$^\prime$. Consequently, the A layer is sandwiched between gapped trivial insulators, and it emerges as a nontrivial QSH layer. This is also consistent with the results of recent work by Liu \textit{et al.}~\cite{liu2011b}. If we consider the AA$^\prime$-AA$^\prime$ sequence as a charge density wave perturbation on the A-A stacked lattice, the stacking fault is equivalent to the domain wall of two charge density waves. This type of domain wall was proposed to accommodate a QSH state~\cite{liu2011b}. In this sense, topological stacking faults in a trivial honeycomb material may offer a platform to realize the QSH effect. In KHgSb, the QSH layer has a large bulk energy gap of approximately 0.2 eV. Therefore, topological edge states can be measured even at the room temperature.  

In conclusion, we predicted the first weak topological insulators by designing odd-layered structures in the honeycomb lattice using \text{ab initio} band structure calculations. The diversity of these ternary compounds allows us to tune the energy gap, the band inversion strength and inter-layer coupling.  The transition from a weak TI to a strong TI, and to a trivial insulator was realized. The most suitable candidate in this family is the single-layered KHgSb, which is a weak TI with large bulk energy gap of 0.24 eV. Even the trivial insulators in this family, i.e. even-layered compounds, can host a quantum spin Hall layer, when a stacking fault occurs in the crystal. The inter-layer difference in real materials can be removed by strong disorders, leading to the existence of weak TIs by forming equivalent single-layered structures. The reported honeycomb lattice can be a prototype to find new weak TIs in layered semiconductors, in which the method of odd-layered structures can be applied. 

\begin{acknowledgments}
We are indebted to Dr. H.J. Zhang and Prof. S.C. Zhang at Stanford University, Prof. J. K\"{u}bler at Technische Universit\"{a}t Darmstadt, Prof. J. Mydosh at Universit\"{a} zu K\"{o}ln and Prof. C.X. Liu at Penn. Stae University  for their great help. B.Y. thanks the  finacial funding by the DFG Project /SPP 1458.
\end{acknowledgments}
%

\end{document}